\documentclass[a4paper]{article}

\usepackage{spconf}
\usepackage[T1]{fontenc}
\usepackage{xcolor}
\usepackage{url}
\usepackage{math_commands}
\usepackage{amsmath}
\usepackage{wasysym}
\usepackage{graphicx}

\title{Dynamic Independent Component Extraction with Blending Mixing Vector:   Lower Bound on Mean Interference-to-Signal Ratio}
\name{Jaroslav \v{C}mejla$^1$, Zbyn\v{e}k Koldovsk\'{y}$^1$, V\'{a}clav Kautsk\'{y}$^1$, and T\"ulay Adali$^2$
\thanks{Part of this work was supported by The Czech Science Foundation through Project No.~20-17720S.}}
\address{$^1$Faculty of Mechatronics, Informatics, 
and Interdisciplinary Studies\\
  Technical University of Liberec, Czech Republic\\
  $^2$Department of Electrical and Computer Engineering\\
  University of Maryland, Baltimore County, Baltimore, MD, 21250, USA
}

\ninept
\begin{document}

\maketitle
\begin{abstract}
    This paper deals with dynamic Blind Source Extraction (BSE) from where the mixing parameters characterizing the position of a source of interest (SOI) are allowed to vary over time. We present a new source extraction model called CvxCSV which is a parameter-reduced modification of the recent Constant Separation Vector (CSV) mixing model. In CvxCSV, the mixing vector evolves as a convex combination of its initial and final values. We derive a lower bound on the achievable mean interference-to-signal ratio (ISR) based on the Cram\'er-Rao theory. The bound reveals advantageous properties of CvxCSV compared with CSV and compared with a sequential BSE based on independent component extraction (ICE). In particular, the achievable ISR by CvxCSV is lower than that by the previous approaches. Moreover, the model requires significantly weaker conditions for identifiability, even when the SOI is Gaussian.
\end{abstract}


\section{Introduction}
Blind Source Separation/Extraction (BSS/BSE) deals with separating a mixture of signals or extracting a signal-of-interest (SOI) from the mixture without prior information or training. It is successfully realized using Independent Component Analysis/Extraction (ICA/ICE) \cite{comon2010handbook}. For example, in audio source separation \cite{ASSSEbook2018}, blind techniques provide attractive alternatives to deep learning methods through direct interpretability \cite{boeddeker2021}. This often happens in problems where the mixture variability is too high so that not all situations can be represented with training examples. 

Even though limiting, so far, research has been focused almost exclusively on static mixing models. 
Recently, there has been a growing interest in dynamic models because real situations involve moving sources. This is typically solved by deploying methods in online processing mode \cite{nakashima2022}.
A recent dynamic (semi-time-variant) mixing model for ICE and Independent Vector Extraction (IVE), called Constant Separating Vector (CSV) addresses this gap \cite{koldovsky2019icassp}. Here, the goal is to estimate a time-invariant beamformer that extracts the source from multiple locations covering the entire motion space \cite{jansky2022}. This way, the parameters are interconnected, and their number is reduced. CSV makes it possible not only to achieve higher accuracy than a fully time-varying model (time-variant mixing and de-mixing parameters) but also to avoid problems caused by the permutation uncertainty of the SOI \cite{koldovsky2021fastdiva}. The fully time-varying model can choose an arbitrary source as the SOI at each movement location, and thus the extracted signal can consist of concatenated intervals from multiple sources.

This paper proposes an effective modification of CSV with a significantly reduced number of parameters, which moderates the demands on the amount of data---lacking in online mode. In the original CSV, the source movement is captured by a time-varying mixing vector whose value is allowed to change from interval to interval. The proposed model enables a continuous or interval-to-interval change of the mixing vector where its current value is a convex combination of the initial and final values. Similar to CSV, the separating vector is assumed to be time-invariant, and hence, the proposed model is called Convex-CSV (CvxCSV). To our best knowledge, this powerful mixing model has not yet been considered in the context of BSE. A similar model was considered by Yeredor et al. in \cite{yeredor2003,weisman2006}; however, it was in the context of BSS, where the entire mixing matrix changes over time, which generally excludes the existence of a constant separating vector.

In this paper, we also derive the Cram\'er-Rao-induced lower bound (CRIB) for the  Interference-to-Signal Ratio (ISR). The bound reveals important features of the model: the minimum ISR achievable by any algorithm based on the model and the identification conditions. The result points to several advantageous properties over the original CSV. First, it is the effective use of the non-stationarity of the mixing model and of the non-stationarity of the SOI (so-called double non-stationarity \cite{koldovsky2022double}) causing  the achievable ISR to be lower than the original CSV \cite{kautsky2020CRLB}. Second, it is the fact that the model is identifiable even when the SOI is circular Gaussian, which is not possible with CSV \cite{kautsky2020CRLB}.

\section{Model}
\subsection{Determined mixture with blending mixing vector}
We consider a complex-valued mixture comprising the SOI and additional noise observed by $d$ sensors.
Let $N$ observed samples be divided into $T$ time intervals of the same length $N_b$, called blocks; in the case that $T=N$ and $N_b=1$, the blocks correspond to the samples. The $n$th observed sample, $n=1,\dots,N$ is described by
\begin{equation}\label{mixing_general}
    \x[n] = \a[t]\s[n] + \y[n],
\end{equation}
where $\s[n]$ and $\y[n]$ stand, respectively, for the SOI and noise; $\a$ stands for the {\em mixing vector} of the SOI, and $t=\lceil\frac{nT}{N}\rceil$; the dependence of $t$ on $n$ will not be explicitly written, for simplicity. 
Let $\a[1]$ and $\a[T]$ be free parameter vectors. The key assumption in CvxCSV is that the SOI movement is a smooth blending of two positions corresponding to $\a[1]$ and $\a[T]$ so that
\begin{equation}\label{blending}
 \a = \lambda_t \a[1] + (1 - \lambda_t) \a[T],
\end{equation}
where $\lambda_t\in[0,1]$, $\lambda_1=1$, $\lambda_T=0$. The case that  $\lambda_t=\frac{t-1}{T-1}$ will be referred to as {\em linear}. 

We now introduce constraints that enable us to write \eqref{mixing_general} in the form of a determined instantaneous mixture; the steps are similar to those in \cite{koldovsky2021fastdiva, kautsky2020CRLB, koldovsky2019TSP}. 

Let $\B= (\g\,\,\, -\gamma_t\mathbf{I}_{d-1})$, where $\gamma_t$ and $\g$ are the upper and lower parts of $\a$, i.e., $\a=(\begin{smallmatrix}\gamma_t\\\g\end{smallmatrix})$; $\mathbf{I}_{d}$ is the $d\times d$ identity matrix. $\B$ is a $(d-1)\times d$ blocking matrix since $\B\a={\bf 0}$. We can introduce $d-1$  background sources $\z[n]=\B\x[n]=\B\y[n]$ since they do not contain the SOI. 
By adopting the assumption of CSV that a {\em time-invariant} separating vector $\w$ exists such that $\s[n]=\w^H\x[n]$ for $n=1,\dots,N$, $\cdot^H$ stands for the conjugate transposition, the de-mixing model with the $d\times d$ de-mixing matrix ${\bf W}_t$ reads
\begin{equation}\label{demixing_matrix}
    \begin{pmatrix}
    \s[n]\\
    \z[n]
    \end{pmatrix} = {\bf W}_t\x[n]=\begin{pmatrix}
    \w^H\\ \B
    \end{pmatrix}\x[n].
\end{equation} 
By inversion, the mixing model is obtained as
\begin{equation}\label{mixing_matrix}
    \x[n]={\bf A}_t
    \begin{pmatrix}
        \s[n]\\
        \z[n]
    \end{pmatrix},
\end{equation}
where ${\bf A}_t={\bf W}_t^{-1}=(\a\,\, \Q)$, 
\begin{equation}
    \Q=\begin{pmatrix}
        {\bf h}^H\\
        \frac{1}{\gamma_t}({\bf g}_{t}{\bf h}^H-\mathbf{I}_{d-1})
    \end{pmatrix},
\end{equation}
and it must be satisfied that $\w^H\a = 1$ for every $t=1,\dots,T$. By \eqref{blending}, this gives two conditions called {\em distortionless constraints}:
\begin{align}\label{distortionconst}
\w^H\a[1] = 1, && \w^H\a[T] = 1.
\end{align}
The other constraint for the equivalence of \eqref{mixing_general} and \eqref{mixing_matrix} is that $\y[n]=\Q\z[n]=\Q\B\x[n]=(\mathbf{I}_d-\a\w^H)\x[n]$, which means that the noise subspace has dimension $d-1$. This restriction pays off for mathematical tractability, and its importance decreases as the number of sensors $d$ increases. 


\subsection{Complexity of CvxCSV vs CSV}\label{sec:cvxcsvvscsv}
The free parameters in CvxCSV are the $d\times 1$ parameter vectors $\a[1]$, $\a[T]$, and $\w$, which are linked through \eqref{distortionconst}. Hence, CvxCSV involves $3d-2$ free parameters.

In CSV \cite{koldovsky2021fastdiva}, each mixing vector $\a[t]$, $t=1,\dots,T$, is a free parameter vector linked with $\w$ through $\w^H\a = 1$. Hence, CSV has $T(d-1)+d$ free parameters. CvxCSV and CSV coincide when $T=2$, and both models correspond to the conventional (static) ICE model \cite{koldovsky2019TSP} when $T=1$.

\subsection{Statistical model}
We consider a similar statistical model of signals as in \cite{koldovsky2021fastdiva,koldovsky2022double}. In particular, $\s$ and $\z$ are assumed to be mutually independent; the samples of $\s$ as well as of $\z$ are assumed to be independently distributed; let their pdfs depend on $t$ and be denoted $p_{s_t}$ and $p_{{\bf z}_t}$, respectively; let $s_t$ and ${\bf z}_t$ symbolize random variables having the corresponding pdfs. By \eqref{demixing_matrix}, the pdf of $\x$, $n=1,\dots,N$, is given by
\begin{equation} 
p_{\mathbf{x}_t}(\x|\a[1],\a[\Tb],\w) = p_{s_t}(\s)p_{\mathbf{z}_t}(\z)|\det(\W)|^2,
\end{equation}
where $\det(\W)=(-1)^{d-1}\gamma_t^{d-2}$; see Eq. (15) in \cite{koldovsky2019TSP}. Hence, the log-likelihood function involving the distortionless constraint (\ref{distortionconst}) reads
\begin{equation}\label{loglik}
\begin{split} 
    \mathcal{L}({\bf x} |\g[1],\g[\Tb],\h) = \sum_{n=1}^N\Bigl\{&\log p_{s_t}(\w^H\x) +\\ 
    + &\log p_{{\bf z}_t}(\B\x) +\\
    + &(d-2)\log|1-\h^H\g[t]|^2\Bigr\},
\end{split}
\end{equation}
where ${\bf x}$ symbolizes the entire observation $\x[1],\dots,\x[N]$.

\section{Cram\'er-Rao-induced bound on ISR}
The Cram\'er-Rao lower bound (CRLB) determines a minimum variance achievable by unbiased estimators of deterministic parameters and corresponds to the asymptotic mean square error of maximum likelihood estimators. 
Similarly to \cite{kautsky2020CRLB}, we use the CRLB to compute the lowest achievable value of the Interference-to-Signal Ratio (ISR) measured at the output of the BSE model proposed in this paper. 

\subsection{ISR}
Let ${\bf w}$, $\widehat{\bf w}$, and $\boldsymbol{\epsilon}$ be, respectively, the true separating vector, its estimate, and an error vector of the asymptotic order $o(1)$. We assume that $\widehat{\bf w}={\bf w}+\boldsymbol{\epsilon}$, thus, $\widehat{\bf w}^H {\bf A}_{t} = ({\bf e}_1 + {\bf A}_{t}^H\boldsymbol{\epsilon})^H$, where ${\bf e}_1$ is the first column of $\mathbf{I}_{d}$.
The estimated SOI then equals
\begin{multline}
    \hat s(n)=\widehat{\bf w}^H\x=\widehat{\bf w}^H{\bf A}_{t}\begin{pmatrix}\s\\ \z\end{pmatrix}
     = \\
    =(1+\epsilon)s(n) + {\bf q}^H{\bf z}(n),
\end{multline}
where ${\bf A}_{t}^H\boldsymbol{\epsilon}=[\epsilon;{\bf q}]$. The ISR measured on the estimated SOI evaluated over all $N$ samples is given by
\begin{equation}\label{eq:ISRk}
   {\tt ISR} =
    \frac{ \left<{\rm E}\left[ \left| {\bf q}^H {\bf z}_{t} \right|^2\right]\right>_t} {|1+\epsilon|^2\left< {\rm E}\left[ \left|  s_{t} \right|^2\right]\right>_t}
    =
    \frac{ {\tt tr}\left[\left< {\bf C}_{{\bf z}_{t}} \right>_t{\bf q}{\bf q}^H\right]} {\left< \sigma^2_{t} \right>_t}+o(1),
\end{equation}
where $\left< \cdot\right>_t$ is the average value of the argument over $t=1,\dots,T$, ${\bf C}_{{\bf z}_{t}}={\rm E}[{\bf z}_{t}{\bf z}_{t}^H]$, and ${\tt tr}[\cdot]$ denotes the trace of the argument. Hence, the approximate lower bound on the mean value of ISR is given by
\begin{equation}\label{eq:ISRdef}
  {\rm E} \left[ {\tt ISR} \right] \apprge
  \frac{ {\tt tr}\left[\left< {\bf C}_{{\bf z}_{t}} \right>_t{\tt cov}[{\bf q}]\right]} {\left<  \sigma^2_{t} \right>_t},
\end{equation} 
where ${\tt cov}[{\bf q}]={\rm E}[{\bf q}{\bf q}^H]$, and the Cram\'er-Rao-induced lower bound (CRIB) on ISR is obtained when ${\tt cov}[{\bf q}]$ is replaced by the CRLB for the parameter vector ${\bf q}$. The important fact is that ${\bf q}$ corresponds to the lower part of $\widehat{\bf w}$, which is $\widehat{\bf h}$ when the true separating and mixing vectors are all unit vectors, that is,  ${\bf w}={\bf a}_{t}={\bf e}_1$. This has been referred to as {\em equivariance}, which significantly simplifies the whole analysis.


\subsection{FIM}
In the CRLB analysis for complex-valued parameters \cite{complCRB}, for any unbiased estimator of a parameter vector $\boldsymbol{\theta}$, 
it holds that
$
{\tt cov}\left[\boldsymbol{\theta}\right] \succeq 
\mathcal{J}^{-1}\left(\boldsymbol{\theta} \right) = 
{\rm CRLB} \left(\boldsymbol{\theta} \right),
$ 
where $\mathcal{J}\left(\boldsymbol{\theta} \right)$ 
is the Fisher Information Matrix (FIM), and ${\bf C}\succeq {\bf D}$ means that 
${\bf C}-{\bf D}$ is a positive semi-definite matrix. The FIM can be 
partitioned as $
\mathcal{J}\left(\boldsymbol{\theta} \right) = \left(\begin{smallmatrix}
 {\bf F} & {\bf P}  \\
 {\bf P}^* &  {\bf F}^*  
  \end{smallmatrix}\right)
$,
where $\cdot^*$ denotes the conjugate value,
  \begin{equation}\label{wirtingerDerivatives}
 {\bf F} =
 {{\rm E}}\left[\frac{\partial \mathcal{L}}{\partial \boldsymbol{\theta}^*}\left(\frac{\partial 
 \mathcal{L}}{\partial
 \boldsymbol{\theta}^*}\right)^H\right], \quad
  {\bf P}  = 
 {{\rm E}}\left[\frac{\partial \mathcal{L}}{\partial \boldsymbol{\theta}^*}\left(\frac{\partial 
 \mathcal{L}}{\partial 
 \boldsymbol{\theta}^*}\right)^T\right], 
  \end{equation}
where the derivatives are defined according to the Wirtinger calculus. 

In our estimation problem, the only free parameters are the lower parts of $\a[1]$, $\a[T]$, and $\w$, i.e., $\g[1]$, $\g[T]$, and $\h$ because of the constraints \eqref{distortionconst}. The $\beta$ in $\w$ can be fixed to $\beta=1$ because this removes a redundant parameter due to the scaling ambiguity in \eqref{mixing_general} ($\a[t]$ and $\s[n]$ can be replaced by $a\a[t]$ and $\s[n]/a$ for any $a\neq 0$) and because the ISR is invariant to the scaling. Therefore, we define the whole parameter vector as $\boldsymbol{\theta}=[{\bf g}_{1}^T, {\bf g}_{T}^T, \h^T]^T$, and hence,
${\bf F}$ and ${\bf P}$ are square matrices of dimension $3(d-1)$. 


\begin{figure*}[ht!]
  \includegraphics[width=\textwidth]{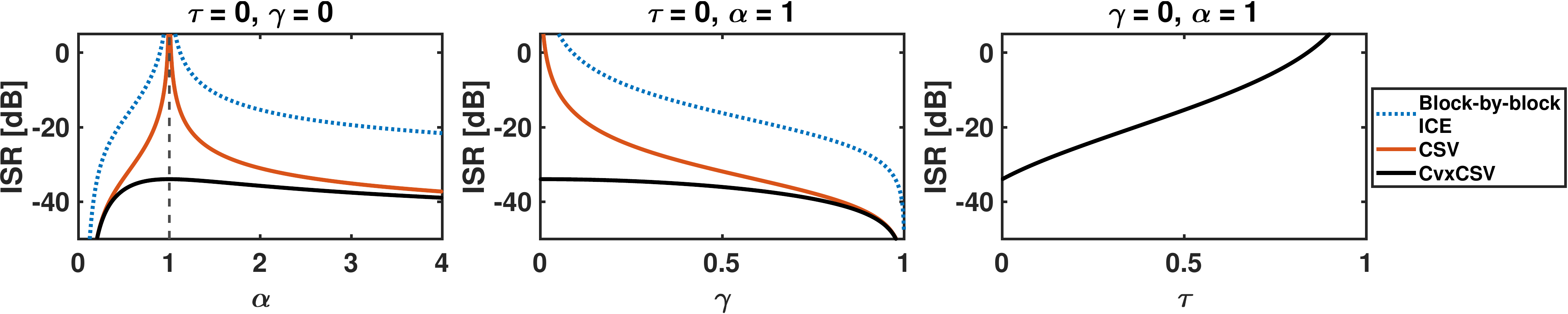}
  \caption{The charts show the theoretically achievable ISR (CRIB) in terms of SOI properties ($\alpha$, $\gamma$ and $\tau$ parameters) and compare the output for block-by-block processing, the CSV model, and the linear variant of the proposed CvxCSV model.}
  \label{fig:results}
\end{figure*}

\subsection{Circular Gaussian background}
From now on, we assume that the background signals $\z[n]$ are circular Gaussian with the covariance matrix $\Cz$, so $\log p_{{\bf z}_t}(\z[n])= -\z[n]^H\Cz^{-1}\z[n]+\text{const.}$ This special case allows a deeper analytical insight and gives an approximate idea of situations where the background is not Gaussian \cite{adali2014SPM,kautsky2020CRLB}. By putting into \eqref{loglik}, the log-likelihood function is equal to
\begin{multline}\label{loglikelihood_gaussian_bcg}
    \mathcal{L}({\bf x}|\boldsymbol{\theta}) =\sum_{n=1}^N \log p_s(\w^H\x) +\\     -  \z[n]^H\Cz^{-1}\z[n] + (d-2)\log|1-\h^H\g[t]|^2.
\end{multline}
The next step is to compute the Wirtinger derivatives \eqref{wirtingerDerivatives} and to evaluate them for $\h=\g={\bf 0}$.
\begin{align}
    \label{nabla_h}
    \nabla^{\h} =&  \left.\frac{\partial\mathcal{L}}{\partial \conj{\h}}\right|_{\h=\mathbf{0}\atop\g[t]=\mathbf{0}} = \sum_{n=1}^N\phi_{t}(s(n))\z[n], \\
    \nabla^{\g[1]} =& \left.\frac{\partial\mathcal{L}}{\partial \conj{\g[1]}}\right|_{\h=\mathbf{0}\atop\g[t]=\mathbf{0}} = -\sum_{n=1}^N(1-\lambda_t)\conj{s(n)}\Cz^{-1}\z[n],  \\
    \label{nabla_gT}
    \nabla^{\g[T]} =& \left.\frac{\partial\mathcal{L}}{\partial \conj{\g[\Tb]}}\right|_{\h=\mathbf{0}\atop\g[t]=\mathbf{0}} = -\sum_{n=1}^N\lambda_t\conj{s(n)}\Cz^{-1}\z[n],
\end{align}
where $\phi_{t}(s) = -\frac{\partial\log p_{s_t}(s,\conj{s})}{\partial s}$ is the score function of the SOI's pdf on the $t$th block. Since the samples are assumed to be independently distributed, the FIM has the structure 
$\mathcal{J}(\boldsymbol{\theta}) = N_b\sum_{t=1}^{\Tb} \mathcal{J}_t(\boldsymbol{\theta})$  
where 
\begin{equation}
    \mathcal{J}_t(\boldsymbol{\theta}) =
    \begin{pmatrix}
        \mathbf{F}_t & \mathbf{P}_t \\
        \conj{\mathbf{P}_t} & \conj{\mathbf{F}_t}
    \end{pmatrix}
\end{equation}
is the FIM corresponding to any sample within the $t$th block where the data are i.i.d. By using \eqref{nabla_h}-\eqref{nabla_gT} for any $n$ and $t=\lceil\frac{nT}{N}\rceil$,
\begin{equation}
\begin{split}
    \mathbf{F}_t = 
    \begin{pmatrix}
    (1-\lambda_{t})^2\mathbf{R}_t & (1-\lambda_{t})\lambda_{t}\mathbf{R}_t &-(1-\lambda_t)\mathbf{I}_{d-1}\\
    (1-\lambda_{t})\lambda_{t}\mathbf{R}_t & \lambda_{t}^2\mathbf{R}_t & -\lambda_{t}\mathbf{I}_{d-1} \\
    -(1-\lambda_t)\mathbf{I}_{d-1} & -\lambda_{t}\mathbf{I}_{d-1} &\kappa_{s_{t}} \Cz
    \end{pmatrix},
\end{split}
\end{equation}
where $\kappa_{s_{t}} = \mathrm{E}[{|\phi_t(s(n))|^2}]$, and $\mathbf{R}_t = \sigma_{t}^2\Cz^{-1}$. Owing to the assumed circularity of $\z[n]$, ${\bf P}_t={\bf 0}$, so $\mathcal{J}(\boldsymbol{\theta})$ is block-diagonal, and we can only focus on the computation of ${\bf F}^{-1}$ to derive the CRLB. The structure of ${\bf F}$ is
\begin{multline}\label{Fstructure}
    \mathbf{F} = \Nb\sum_{t=1}^\Tb \mathbf{F}_t = \Nb \left(\begin{array}{c|c}
    A & B \\ \hline
    C & D
    \end{array}\right) =\\
    \Nb
    \left(\begin{array}{cc|c}
    \sum\limits_{t=1}^{\Tb}(1-\lambda_{t})^2 \mathbf{R}_t & \sum\limits_{t=1}^{\Tb}(1-\lambda_{t})\lambda_{t}\mathbf{R}_t & -\frac{\Tb}{2}\mathbf{I}_{d-1}\\
    \sum\limits_{t=1}^{\Tb}(1-\lambda_{t})\lambda_{t}\mathbf{R}_t & \sum\limits_{t=1}^{\Tb}\lambda_{t}^2 \mathbf{R}_t &  -\frac{\Tb}{2}\mathbf{I}_{d-1} \\ \hline
    -\frac{\Tb}{2}\mathbf{I}_{d-1} & -\frac{\Tb}{2}\mathbf{I}_{d-1} & \sum\limits_{t=1}^{\Tb}\kappa_{s_{t}} \Cz
    \end{array}\right).
\end{multline}
By the block-matrix inversion lemma,
\begin{equation}\label{blocklemma}
    \mathbf{F}^{-1} = \left[\Nb\begin{pmatrix}
    A & B \\
    C & D
    \end{pmatrix}\right]^{-1}= \frac{1}{\Nb}\begin{pmatrix}
    I & J\\
    K & L
    \end{pmatrix},
\end{equation}
and the CRLB for the lower part of ${\bf w}$, which is ${\bf h}$, corresponds to the lower corner of ${\bf F}^{-1}$, which is
\begin{equation}
    \mathrm{CRLB}(\h)|_{\h[t]=\mathbf{0}\atop\g[t]=\mathbf{0}} = \frac{1}{\Nb} L = \frac{1}{\Nb} (D-CA^{-1}B)^{-1}.
\end{equation}
By putting \eqref{Fstructure} into \eqref{blocklemma} and then into \eqref{eq:ISRdef}, the desired lower bound on ISR is obtained. However, the closed-form expression cannot be derived in general. We now briefly discuss two special cases that enable deeper analytical insights.

First, as discussed in Section~\ref{sec:cvxcsvvscsv}, the CvxCSV model coincides with CSV for $T=2$ where the corresponding bounds must be the same. This is easily seen since when $\lambda_1=1$ and $\lambda_2=0$, the block $A$ in \eqref{Fstructure} becomes block-diagonal, and the FIM corresponds with Equation (67) in \cite{kautsky2020CRLB}.

Second, the closed-form bound on ISR can be expressed for the special case when ${\bf R}_t$ is independent of $t$ and when we have the linear blending case, i.e., $\lambda_t=\frac{t-1}{T-1}$. The computations are straightforward, however, we do not show them here due to limited space. The resulting bound for such a special case says that
\begin{equation}
    {\rm E} \left[ {\tt ISR} \right] \apprge
    \frac{1}{N}\frac{d-1}{\left<\overline{\kappa}_{s_t}\right>_t-1},
\end{equation}
where $\overline{\kappa}_{s_t}=\kappa_{s_t}\sigma_{s_t}^2$ and it holds that $\overline{\kappa}_{s_t}\geq 1$ and $\overline{\kappa}_{s_t}=1$ if and only if the SOI is circular Gaussian within the $t$th block. This result is similar to Equation (78) in \cite{kautsky2020CRLB}, which corresponds to CSV when the SOI has the same distribution on all blocks. The result points to the important strength of both dynamic models: The lower bound for ISR is proportional to $N^{-1}$ and not to $N_b^{-1}$. For CSV, the SOI must {\em not} be circular Gaussian in this special case, otherwise, the model is not identifiable (the bound is infinite). For CxvCSV, this condition is relaxed: for example, it suffices that the SOI is non-Gaussian within, at least, one block. 

When the SOI is non-stationary, the identifiability conditions under CxvCSV are even more relaxed than under CSV, allowing the SOI to be Gaussian throughout the observation period. We demonstrate this fact by a numerical comparison of the bounds in the following section.

\section{Numerical evaluation}
In this section, we evaluate the CRIB for specific scenarios involving (non-)stationary, (non-)Gaussian, and (non-)circular SOI to analyze the influence of these signal features on the achievable extraction accuracy. We compare the results with similar CRIBs for the previous dynamic BSE models. Namely, we consider a naive blind approach where the conventional ICE is applied block-by-block to adapt to dynamic mixing; the approach is denoted as BICE in \cite{kautsky2020CRLB}. Next, the comparison includes the original CSV model from \cite{koldovsky2021fastdiva}.



The CRIBs are evaluated for cases where the number of sensors is $d = 5$ and the number of observed samples is $N = 5000$. The data are assumed to be divided into $T = 10$ non-overlapping blocks of the same length, hence, $N_b=500$. The SOI is assumed to be distributed according to the complex-valued Generalized Gaussian distribution (GGD) \cite{cGGD}. The pdf corresponding to a normalized GGD random variable (zero mean and unit variance) is given by \cite{GGD_Loesch_yang}
\begin{equation}\label{ggdpdf}
    p(s,\conj{s}) = \frac   {
                            \alpha\rho\exp(- [ \frac{\rho/2}{\gamma^2 - 1} (\gamma s^2 + 
                            \gamma(\conj{s})^2-2s\conj{s}) ]^\alpha )
                            }
                            {
                            \pi\Gamma(1/\alpha)(1-\gamma^2)^\frac{1}{2}
                            },
\end{equation}
where $\rho = \frac{\Gamma(2/\alpha)}{\Gamma(1/\alpha)}$; $\Gamma(\cdot)$ denotes the Gamma function; $\alpha\in[0,\infty]$ is a shape parameter; $\gamma\in[0,1]$ controls the level of (non-)circularity. Specifically, the distribution is super-Gaussian for $\alpha<1$, Gaussian for $\alpha=1$, and sub-gaussian for $\alpha>1$; it is circular when $\gamma=0$ and non-circular  for $\gamma>0$; it is degenerate when $\gamma=1$. The corresponding score function is equal to
\begin{equation}\label{ggdscore}
    \phi(s,\conj{s}) = \frac    {
                                    2\alpha(\rho/2)^\alpha
                                }
                                {
                                    (\gamma^2-1)^\alpha
                                }
                        (\gamma s^2 + \gamma(\conj{s})^2 - 2s\conj{s})^{\alpha-1} 
                        (\gamma s - \conj{s}).
\end{equation}
By definition, for a SOI whose pdf on the $t$th block is GGD and its variance is $\sigma^2_t$, it holds that
\begin{equation}
    \overline{\kappa}_{s_t} = \kappa_{s_t}\sigma_{s_t}^2 = \frac{\alpha^2\Gamma(2/\alpha)}{(1-\gamma^2)\Gamma^2(1/\alpha)}.
\end{equation}
To simulate the (non-)stationarity of the SOI, we consider its variance on the $t$th block to be given by  
\begin{equation}
    \sigma_{s_t} = \tau + (1-\tau)\sin\left(\frac{\pi t}{2T}\right),
\end{equation}
where $\tau \in [0,1]$ controls the level of non-stationarity. For $\tau = 1$, the SOI is stationary (i.i.d.) throughout the observation period; otherwise, it is non-stationary. The background sources ${\bf z}_t$ are assumed to be circular Gaussian, i.e., $\gamma = 0$, $\alpha=1$, with identity covariance, i.e., $\Cz[t]=\mathbf{I}_{d-1}$.

The CRIBs for all three compared models are shown in Fig.~\ref{fig:results}. The left-hand chart of Fig.~\ref{fig:results} shows the CRIB for circular non-stationary SOI ($\gamma=0$, $\tau=0$) in terms of varying  $\alpha$. It can be noticed that the SOI's non-stationarity ensures that the CvxCSV model is identifiable even in the case of circular Gaussian SOI ($\alpha=1$). In contrast, the other models are unidentified (the CRIB grows to infinity). The second chart of Fig.~\ref{fig:results} depicts the results for non-stationary Gaussian SOI ($\tau=0$, $\alpha=1$) in the case of varying circularity parameter. The CRIBs of all three models decrease as the SOI becomes more non-circular ($\gamma \to 1$). Finally, the right-hand chart in Fig.~\ref{fig:results} shows the results for circular Gaussian SOI ($\gamma = 0$, $\alpha = 1$) when its level of (non-)stationarity is changing. In this case, the CSV and the block-by-block ICE models are not identifiable; therefore, the only displayed result is for CvxCSV. The corresponding CRIB decreases with increasing non-stationarity of the SOI ($\tau \to 0$), and vice versa. For $\tau = 1$, that is, when the SOI is stationary and circular Gaussian, the model is not identifiable. 

\section{Conclusion}
We have proposed a new BSE model for extracting a SOI whose position given by the mixing vector changes in time. The change is approximated by a convex combination of the initial and final values of the vector, thereby reducing the number of free parameters. The analysis of the achievable accuracy in terms of mean ISR shows that with CvxCSV a higher extraction accuracy can be achieved than with the original CSV model or the naive sequential ICE. At the same time, the model enables the identification of a wider set of signals, especially Gaussian signals.

Future research will be focused on the development of efficient algorithms that achieve the optimum accuracy given by the bound, as has been achieved with the other BSE models \cite{koldovsky2021fastdiva,kautsky2020CRLB}. Algorithms based on the CvxCSV model have potential applications in situations where very little data is available. For example, this occurs when the SOI moves rapidly so that the change in its position must be estimated from short observations obtained through a correspondingly short time interval.




 
\bibliographystyle{IEEEtran}


\begin{thebibliography}{10}
\providecommand{\url}[1]{#1}
\csname url@samestyle\endcsname
\providecommand{\newblock}{\relax}
\providecommand{\bibinfo}[2]{#2}
\providecommand{\BIBentrySTDinterwordspacing}{\spaceskip=0pt\relax}
\providecommand{\BIBentryALTinterwordstretchfactor}{4}
\providecommand{\BIBentryALTinterwordspacing}{\spaceskip=\fontdimen2\font plus
\BIBentryALTinterwordstretchfactor\fontdimen3\font minus
  \fontdimen4\font\relax}
\providecommand{\BIBforeignlanguage}[2]{{%
\expandafter\ifx\csname l@#1\endcsname\relax
\typeout{** WARNING: IEEEtran.bst: No hyphenation pattern has been}%
\typeout{** loaded for the language `#1'. Using the pattern for}%
\typeout{** the default language instead.}%
\else
\language=\csname l@#1\endcsname
\fi
#2}}
\providecommand{\BIBdecl}{\relax}
\BIBdecl

\bibitem{comon2010handbook}
P.~Comon and C.~Jutten, \emph{Handbook of Blind Source Separation: Independent
  Component Analysis and Applications}, ser. Independent Component Analysis and
  Applications Series.\hskip 1em plus 0.5em minus 0.4em\relax Elsevier Science,
  2010.

\bibitem{ASSSEbook2018}
E.~Vincent, T.~Virtanen, and S.~Gannot, \emph{Audio Source Separation and
  Speech Enhancement}, 1st~ed.\hskip 1em plus 0.5em minus 0.4em\relax Wiley
  Publishing, 2018.

\bibitem{boeddeker2021}
\BIBentryALTinterwordspacing
C.~Boeddeker, F.~Rautenberg, and R.~Haeb-Umbach, ``A comparison and combination
  of unsupervised blind source separation techniques,'' 2021. [Online].
  Available: \url{https://arxiv.org/abs/2106.05627}
\BIBentrySTDinterwordspacing

\bibitem{nakashima2022}
\BIBentryALTinterwordspacing
T.~Nakashima and N.~Ono, ``Inverse-free online independent vector analysis with
  flexible iterative source steering,'' 2022. [Online]. Available:
  \url{https://arxiv.org/abs/2209.00937}
\BIBentrySTDinterwordspacing

\bibitem{koldovsky2019icassp}
Z.~Koldovsk\'y, J.~M\'alek, and J.~Jansk\'y, ``Extraction of independent vector
  component from underdetermined mixtures through block-wise determined
  modeling,'' vol. 7903--7907, May 2019.

\bibitem{jansky2022}
J.~Jansk{\'y}, Z.~Koldovsk{\'y}, J.~M{\'a}lek, T.~Kounovsk{\'y}, and
  J.~{\v{C}}mejla, ``Auxiliary function-based algorithm for blind extraction of
  a moving speaker,'' \emph{EURASIP Journal on Audio, Speech, and Music
  Processing}, vol. 2022, no.~1, p.~1, Jan 2022.

\bibitem{koldovsky2021fastdiva}
Z.~Koldovský, V.~Kautský, P.~Tichavský, J.~Čmejla, and J.~Málek, ``Dynamic
  independent component/vector analysis: Time-variant linear mixtures separable
  by time-invariant beamformers,'' \emph{IEEE Transactions on Signal
  Processing}, vol.~69, pp. 2158--2173, 2021.

\bibitem{yeredor2003}
A.~Yeredor, ``{TV-SOBI}: An expansion of sobi for linearly time-varying
  mixtures,'' in \emph{Proceedings of The 4th International Symposium on
  Independent Component Analysis and Blind Source Separation (ICA2003)}, April
  2003.

\bibitem{weisman2006}
T.~Weisman and A.~Yeredor, ``Separation of periodically time-varying mixtures
  using second-order statistics,'' in \emph{Independent Component Analysis and
  Blind Signal Separation}, J.~Rosca, D.~Erdogmus, J.~C. Pr{\'i}ncipe, and
  S.~Haykin, Eds.\hskip 1em plus 0.5em minus 0.4em\relax Berlin, Heidelberg:
  Springer Berlin Heidelberg, 2006, pp. 278--285.

\bibitem{koldovsky2022double}
Z.~Koldovský, V.~Kautský, and P.~Tichavský, ``Double nonstationarity: Blind
  extraction of independent nonstationary vector/component from nonstationary
  mixtures---{Algorithms},'' \emph{IEEE Transactions on Signal Processing},
  vol.~70, pp. 5102--5116, 2022.

\bibitem{kautsky2020CRLB}
V.~{Kautský}, Z.~{Koldovský}, P.~{Tichavský}, and V.~{Zarzoso},
  ``{Cramér–Rao} bounds for complex-valued independent component extraction:
  Determined and piecewise determined mixing models,'' \emph{IEEE Transactions
  on Signal Processing}, vol.~68, pp. 5230--5243, 2020.

\bibitem{koldovsky2019TSP}
Z.~Koldovsk\'y and P.~Tichavsk\'y, ``Gradient algorithms for complex
  non-{Gaussian} independent component/vector extraction, question of
  convergence,'' \emph{IEEE Transactions on Signal Processing}, vol.~67, no.~4,
  pp. 1050--1064, Feb 2019.

\bibitem{complCRB}
T.~Menni, E.~Chaumette, P.~Larzabal, and J.~P. Barbot, ``New results on
  deterministic {Cram\' er--Rao} bounds for real and complex parameters,'' in
  \emph{IEEE Trans. Signal Processing}, vol.~60, March 2012, pp. 1032--1049.

\bibitem{adali2014SPM}
T.~Adal{\i} and P.~J. Schreier, ``Optimization and estimation of complex-valued
  signals: Theory and applications in filtering and blind source separation,''
  \emph{IEEE Signal Processing Magazine}, vol.~31, no.~5, pp. 112--128, 2014.

\bibitem{cGGD}
M.~Novey, T.~Adal{\i}, and A.~Roy, ``A complex generalized {Gaussian}
  distribution--characterization, generation, and estimation,'' \emph{IEEE
  Trans. Signal Processing}, vol.~58, pp. 1427--1433, March 2010.

\bibitem{GGD_Loesch_yang}
B.~Loesch and B.~Yang, ``{Cramér-Rao} bound for circular and noncircular
  complex independent component analysis,'' \emph{IEEE Transactions on Signal
  Processing}, vol.~61, no.~2, pp. 365--379, 2013.

\end{thebibliography}

\end{document}